\documentclass{jaa}
\usepackage{natbib}
\usepackage[flushleft]{threeparttable}
\usepackage{amsmath}
\usepackage{xcolor}
\usepackage[colorlinks]{hyperref}

\usepackage{graphicx}

\begin{document}\sloppy

\title{Temporal and spectral study of the X-ray pulsar 2S 1553–542 during the 2021 outburst}

\author{Manoj Mandal\textsuperscript{1} and Sabyasachi Pal\textsuperscript{1,*}}
\affilOne{\textsuperscript{1}Midnapore City College, Kuturia, Bhadutala, 721129, India.\\}


\twocolumn[{

\maketitle

\corres{sabya.pal@gmail.com}


\begin{abstract}
We study the timing and spectral properties of the X-ray pulsar 2S 1553–542 using the {\it NuSTAR}, and {\it NICER} during the outburst in January–February 2021. During the outburst, the spin period of the neutron star was $P\sim 9.2822$ s based on {\it NuSTAR} data. 
The pulse profiles are studied using different NICER observations, which implies that the profile is more or less sinusoidal with a single peak and  the beaming patterns are mostly dominated by the pencil beam. 
The {\it NICER} spectra of the source are studied for different days of the outburst and can be well described by a model consisting of a blackbody emission and power law along with a photoelectric absorption component. The variation of spectral parameters with luminosity is studied over the outburst. The photon index shows anti-correlation with luminosity below the critical luminosity, which implies that the source was accreting in the sub-critical accretion regime during the {\it NICER} observations. We also report the anti-correlation between pulsed fraction (PF) and luminosity of the 2S 1553–542 using {\it NICER} observations.  The evolution of spin-up rate with luminosity is studied during the outburst, which implies that both are strongly correlated. The torque-luminosity model is applied to estimate the magnetic field at different spin-up rates. The magnetic field is estimated to be $\simeq 2.56  \times 10^{12}$ G from the torque-luminosity model using the source distance of 20 kpc. The magnetic field is also estimated using the critical luminosity, which is also consistent with our findings.
\end{abstract}


}]


\doinum{12.3456/s78910-011-012-3}
\artcitid{\#\#\#\#}
\volnum{000}
\year{0000}
\pgrange{1--}
\setcounter{page}{1}
\lp{1}

\section{Introduction}

\label{intro}
The X-ray transient 2S 1553–542 was discovered using Small Astronomy Satellite 3 (SAS-3) in 1975 \citep{Ap78} during the Galactic plane survey. A strong pulsation with the period of 9.3 was found \citep{Ke83} with an orbital period P$_{orb}$ of nearly 30 d.
During the outburst in 2007-2008, the spectral and timing properties of the X-ray pulsar were studied using the  Rossi X-ray Timing Explorer ({\it RXTE}) citep{Pa12}. The {\it RXTE} allowed us to improve the orbital parameters of the system and to trace the spectral evolution in the energy range of 2.5–30 keV. During the outburst, the spin period $P_{spin}$ was 9.2829$\pm$0.0003 s with period derivative $\dot{P}$ $\sim$10$^{-9}$ s s$^{-1}$ and the pulsed fraction showed a negative correlation with the time of the outburst. The pulse profile was single-peaked and featureless, and variability in the pulse profile was observed during the outburst. An energy-dependent pulse profile was also studied, but no significant variation in the pulse profile with energy was found. Significant variability of the pulse fraction was observed in {\it RXTE}/PCA observations, and PF was decreased with the decay of the flaring activity. The power density spectrum showed an interesting feature, the fundamental line with its five harmonics was visible. The variation of different spectral parameters with different orbital phases was studied, and it was shown that the source hard X-ray spectra in all available intensity states can be well explained with the combination of a broken power law and a blackbody component (with temperature varying between 2.5 and 4 keV) \citep{Pa12}.

The X-ray pulsar 2S 15553--542 went through another outburst in 2015. During this outburst, the temporal and spectral properties of the source were studied using Chandra and {\it NuSTAR} data \citep{Ts16}. Based on the {\it Fermi}/GBM data, the orbital parameters of the system were substantially improved, which allowed determining the spin period of the neutron star $P$ = 9.27880(3) s and a local spin-up $\dot{P}$ = --7.5$\times10^{-10}$ s s$^{-1}$ due to the mass accretion during the {\it NuSTAR} observations. Assuming accretion from the disc and using standard torque models, the distance to the system was estimated as d = 20$\pm$4 kpc. From the timing analysis, the single-peak-shaped pulse profile was found with a barely noticeable dependence on energy, which is quite similar to the previous results \citep{Pa12}. The cyclotron absorption feature near 23.5 keV was found during the declining stage of the outburst with a corresponding magnetic field strength $\sim 3 \times 10^{12}$ G \citep{Ts16}. During the 2021 outburst, the energy spectrum was studied using {\it NuSTAR} and the luminosity dependence of the cyclotron line was investigated \citep{Mal22}. A state transition is reported above the critical luminosity of $4\times10^{37}$ erg s$^{-1}$ \citep{Mal22}.

The X-ray pulsar went through an outburst after a quiescence of nearly six years as detected by Burst Alert Telescope (BAT) onboard {\it Swift} and Gas Slit Camera (GSC) onboard {\it MAXI} on January 2021 \citep{Ma21}. The X-ray flux started to increase in early January of 2021, and the duration of the outburst was nearly six weeks. We study different temporal and spectral properties of the X-ray pulsar in the soft X-ray band using the Neutron star Interior Composition Explorer ({\it NICER}) and in the hard X-ray band using the Nuclear Spectroscopic Telescope Array ({\it NuSTAR}). We also look for the evolution of the timing and spectral properties in the range of 0.3--79 keV during the outburst in 2021 using  {\it NICER}, and {\it NuSTAR} data. The emission mechanism, beaming patterns, and accretion regime is investigated using timing and spectral study during the outburst.

We have studied the variation of photon index and pulse fraction with luminosity during the outburst. We also look at the evolution of the pulse profile, spin-up rates, and beaming patterns of the source at different luminosities. 
 We look for any correlation between the spin-up rates and luminosity for the source during the outburst. The torque-luminosity model is used to estimate the magnetic dipole moment and magnetic field from the spin-up rates and luminosity correlation.

The data reduction and analysis methods are discussed in Section \ref{obs}. We have summarized the results of the current study in Section \ref{res}. The discussion and conclusion are summarized in Section \ref{dis} and \ref{con} respectively.

\begin{figure}[t]
\centering{
\includegraphics[width=8.8cm]{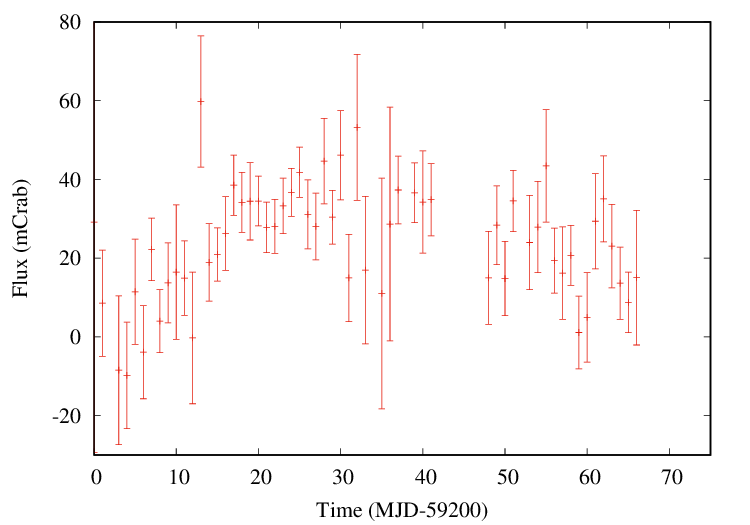}
	\caption{ {\it Swift}/BAT (15--50 keV) detects an outburst form 2S 1553--542 during January-February 2021.
	}
\label{fig:BAT}}
\end{figure}

\section{Observation and data analysis}
\label{obs}
 An outburst from the X-ray pulsar 2S 1553--542 was detected in 2021 and followed up using different satellites in multi-wavelength regions. The Burst Alert Telescope ({\it BAT}) onboard The Neil Gehrels {\it Swift} Observatory \citep{Ge04} is sensitive to hard X-ray (15--50 keV) \citep{Kr13}. To study the evolution of the outburst, we utilize data from various all-sky X-ray monitors, including  {\it Swift}/BAT (15--50 keV), {\it MAXI}/GSC (2--20 keV), {\it Fermi}/GBM (12--25 keV). {\it NuSTAR} data taken at the peak of the 2021 outburst as well as {\it NICER} data were taken on various days of the outburst. We used  {\tt HEASOFT} version 6.28 for the data reduction and analysis. 
 
 We have used the results of the {\it BAT} transient monitor during the outburst, which were provided by the {\it BAT} team. The flux peaked in the third week of January and continued for around six weeks as seen  by the {\it Swift}/BAT. 
 
In the present study, we used the measurements of spin frequency, and 12--25 keV pulsed flux measurements with the {\it Fermi}/GBM \citep{Ca09, Fi09}. The {\it Fermi}/GBM also detected the outburst from 2S 1553--542, which continued for almost six weeks with a peak pulsed flux of $\sim$0.24 keV cm$^{-2}$ s$^{-1}$ on MJD 59240. 

\begin{table*}
\centering
\caption{Log of {\it NuSTAR}, and {\it NICER}  observations.} 
\label{tab:log_table}
\begin{tabular}{|c|c|c|c|c|} 
	\hline
	Instrument & Start time       & Date & Exposure  & Obs. ID       \\
		   &   (MJD)          & (yyyy-mm-dd)& (ks) &                   \\
\hline
{\it NuSTAR} & 59236.95 & 2021-01-22 & 28.3 & 90701302002\\

\hline

          & 59268.08 & 2021-02-23  & ~1.4 & 3202030101 (Obs 1)\\
{\it NICER}	  & 59271.62 & 2021-02-26  & ~2.0 & 3202030102 (Obs 2)\\
	  & 59272.02 & 2021-02-27  & ~0.4 & 3202030103 (Obs 3)\\
          & 59274.02 & 2021-03-01  & ~0.3 & 3202030104 (Obs 4)\\
\hline
\end{tabular}
\end{table*}
 
\subsection{{\it NuSTAR} observation}
The Nuclear Spectroscopic Telescope Array ({\it NuSTAR}) observatory comes with two co-aligned, identical X-ray telescope systems operating in a wide energy range of 3–79 keV. Separate solid-state CdZnTe pixel detector systems in each telescope usually referred to as focal plane modules A and B (FPMA and FPMB; \citep{Ha13}), have a spectral resolution of 400 eV at 10 keV and 900 eV at 68 keV (FWHM), respectively. {\it NuSTAR} performed an observation of 2S 15553--542 close to the peak of the outburst (MJD 59236.95) with a total exposure of 28.35 ks. The observation log of {\it NuSTAR} is shown in Table \ref{tab:log_table}. The data is reduced using the {\tt  NuSTARDAS pipeline} provided under {\tt HEASOFT} with the latest {\tt caldb} and we defined the source and background using {\tt ds9}. The light curve and spectra of the source and background are generated from circular regions centering the source with radii of 50 arcsec and 100 arcsec using {\tt nuproducts} scripts provided by the {\tt NuSTARDAS pipeline}. The light curve has been extracted using the science event data in different energy ranges with bin sizes of 0.1 s for both the modules (FPMA/FPMB) of {\it NuSTAR}.  The {\tt efsearch} task in {\tt FTOOLS} is used to find out the best period in the time series of the barycenter and background corrected data sets. The folding method of the light curve over a trial period to get the best period by $\chi{^2}$ maximizing process \citep{Le87} over 32 phase bins in each period is used. After getting the best spin period, pulse profiles are obtained using the {\tt efold} task in {\tt FTOOLS} by folding light curves with the best spin period. 

\subsection{{\it NICER} observation}
The Neutron Star Interior Composition Explorer ({\it NICER}) onboard the International Space Station is a non-imaging, soft X-ray telescope. The main part of {\it NICER} is the X-ray Timing Instrument (XTI), which operates in a soft X-ray region (0.2--12keV) \citep{Ge16}. {\it NICER} observed the source for one week during the declining phase of the outburst. Table \ref{tab:log_table} summarizes the log of {\it NICER} observations used for the current study. The {\it NICER} data has been processed with {\tt  NICERDAS} in {\tt HEASOFT}. We have created clean event files by applying the standard calibration and filtering tool {\tt nicerl2} to the unfiltered data. We have extracted light curves for different energy ranges with a bin size of 0.1 s and spectra using {\tt XSELECT}. The task {\tt barycorr} is used to apply barycentric corrections for timing analysis. 

We have fitted the NICER spectra in {\tt XSPEC} with the redistribution matrix file (RMF) and {\it NICER} ancillary response file (ARF) provided by {\it NICER} team\footnote{\url{https://heasarc.gsfc.nasa.gov/docs/nicer/proposals/nicer_tools.html}}. We also extracted {\it NICER} spectra for different days of the outburst during the declining phase.
The good time intervals are selected for the timing analysis according to the following criteria: The ISS was not in the South Atlantic anomaly (SAA) zone, the source elevation was $>$ 20$^\circ$ above the Earth limb, and the source direction was at least 30$^\circ$ from the bright Earth.
The background corresponding to each epoch of the observation was simulated by using the {\tt nibackgen3C50}\footnote{\url{https://heasarc.gsfc.nasa.gov/docs/nicer/tools/nicer_bkg_est_tools.html}} tool \citep{Re22}.



 \begin{figure*}[t]
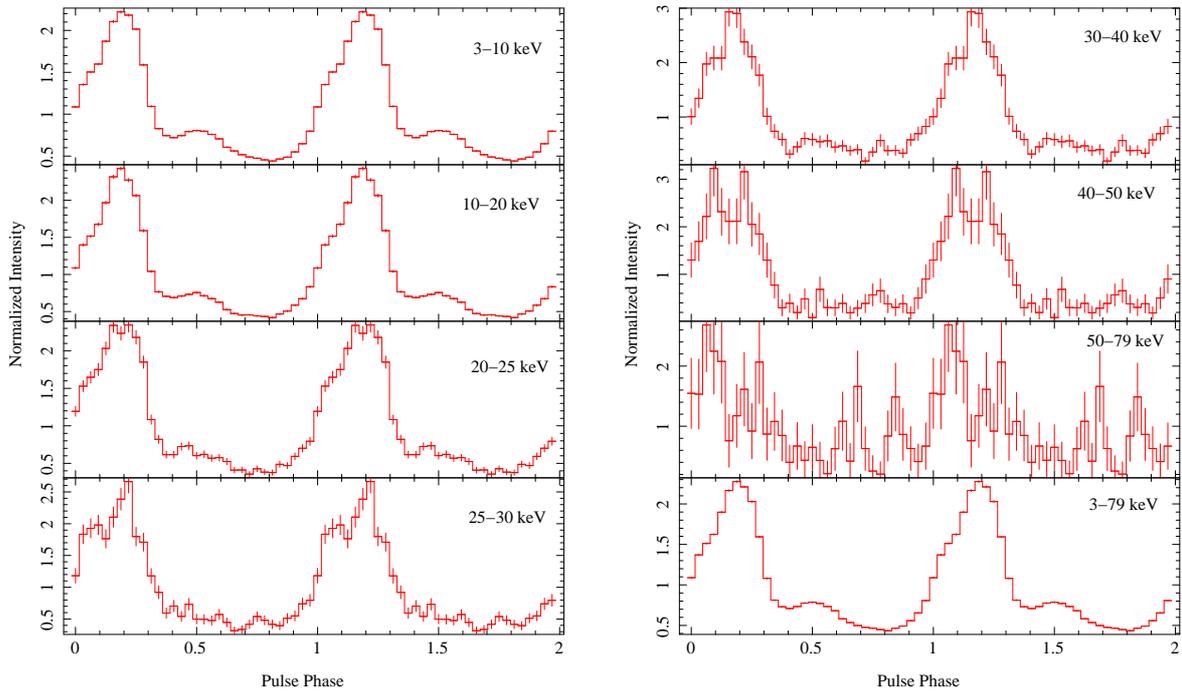

\centering{
\includegraphics[width=8.0cm]{erp2_nu2002.eps}
\includegraphics[width=8.0cm]{erp1_nu2002.eps}
\caption{Energy dependent pulse profiles of 2S 15553--542 using data from {\it NuSTAR} observation during the 2021 outburst. 
}
\label{fig:energy_dependent_pulse_profile}
}
\end{figure*}

\begin{figure}
\centering{
\includegraphics[width=8.8cm]{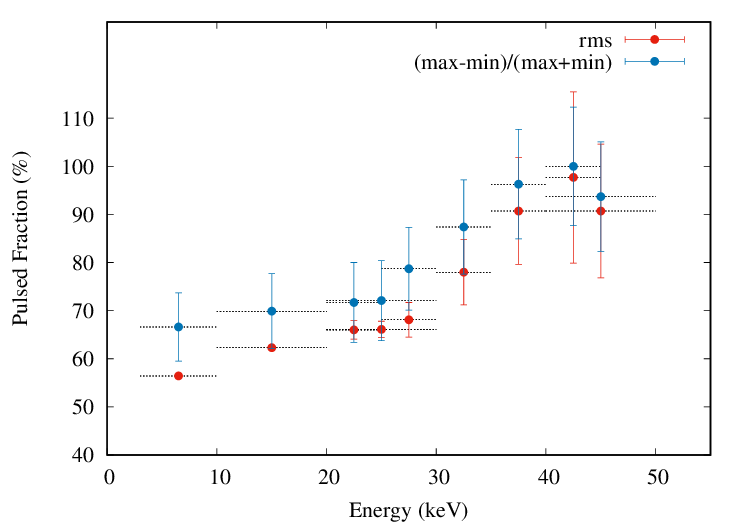}
\caption{Evolution of pulse fraction with energy using {\it NuSTAR} observation. The horizontal error bars indicate the energy range for which PF is estimated and the vertical error bars indicate the errors of corresponding measurements respectively. The figure shows local maxima near cyclotron line energy $\sim$27 keV.}
\label{fig:pulse_fraction}}
\end{figure}

\section{Results}
\label{res}
The X-ray pulsar 2S 15553--542 went through an outburst during January-February 2021, detected by {\it Fermi}/GBM, {\it Swift}/BAT\footnote{\url{https://swift.gsfc.nasa.gov/results/transients/}} and {\it MAXI}/GSC which reached a maximum flux during the last week of January 2021. We used \textit{MAXI} \citep{Ma09} final data products (light curves) as well as the \textit{Fermi} \citep{Fi09, Me09} pulse frequencies and pulsed flux evolution data.
Figure \ref{fig:BAT} shows the variation of flux during the outburst using {\it Swift}/BAT (15--50 keV). The total duration of the outburst was around 6 weeks, which started in the first week of January 2021 and continued till the second week of February 2021. We have summarized the results of the timing and spectral analysis of 2S 15553--542 during the recent outburst in 2021 using data from {\it NuSTAR} and {\it NICER}.

\subsection{Evolution of timing parameters}
We have investigated the evolution of the different timing parameters during the outburst using {\it NuSTAR}, and {\it NICER} observations. The spin period of the pulsar during the outburst was $P$ = $9.2822\pm0.0001$ s using {\it NuSTAR} data, which is consistent with the pulse period evolution history as recorded with {\it Fermi}/GBM during the outburst\footnote{\url{https://gammaray.nsstc.nasa.gov/gbm/science/pulsars}}. 

We also look at the energy dependence of pulse profiles as well as the temporal variation of pulse profiles during the outburst. Figure \ref{fig:energy_dependent_pulse_profile} shows the energy-dependent pulse profiles for different energy ranges of {\it NuSTAR} observation. The variation of the pulse profile over different energy bands is shown in Figure \ref{fig:energy_dependent_pulse_profile}.
The pulse profile shows single-peaked a nearly sinusoidal nature. The pulse profile shows an evolution near the fundamental cyclotron line energy $\sim$27 keV \citep{Mal22}. As the scattering cross-section is changed near CRSF, the emission mechanism, and the beaming pattern are affected. Figure \ref{fig:energy_dependent_pulse_profile} also shows that in the energy ranges of 25--30 keV, the pulse profile changes, which corresponds to the CRSF line energy. 

The variation of Pulse Fraction (PF) with energy is studied during the {\it NuSTAR} observation. PF can be defined as the ratio between the difference of maximum intensity ($I_{max}$) and minimum intensity ($I_{min}$) to their sum: [$(I_{max} - I_{min})/(I_{max} + I_{min})$]. We have also calculated the RMS pulse fraction using the following formula,
\begin{equation}
    PF = \frac{1}{\sqrt{N}} \left[\sum \limits_{i=1}^N (p_{i} -\Bar{p})^2\right]^\frac{1}{2}
\end{equation}
where $N$ is the total number of phase bins, $p_{i}$ is the count rate in the {\it i}th phase bin of the pulse profile, and $\Bar{p}$ is the average count rate. Figure \ref{fig:pulse_fraction} shows the variation of PF for different energy ranges for which the energy-resolved pulse profile is studied, and the horizontal bars represent the energy ranges for which the PF is calculated. The PF shows a positive correlation with energy. We have found that the RMS pulse fraction increases from $\sim$55\% (3--10 keV) to $\sim$90\% (40--50 keV) during the {\it NuSTAR} observation. 

Figure \ref{fig:pulse_fraction} shows the variation of pulse fraction with energy using {\it NuSTAR} data.  
The Figure \ref{fig:pulse_fraction} shows local maxima near $\sim27$ keV which is close to the CRSF fundamental line. The scattering cross-section changes near the CRSF, which affects the beaming pattern and pulse profile. 
We have found a local feature in the pulsed fraction near the cyclotron line at 27 keV and the pulse profile (25--30 keV) also showed an evolution near this energy. The beaming pattern may be affected near the cyclotron line energy, which has an impact on timing parameters.

\begin{figure}
\centering{
\includegraphics[width=8.0cm]{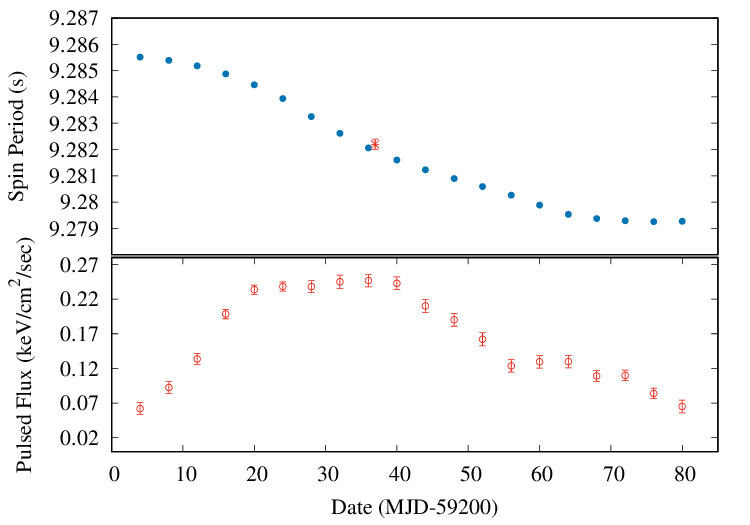}
	\caption{The variation of the spin period during the outburst using {\it Fermi}/GBM is shown with blue circles. The red asterisks show the pulse period calculated from {\it NuSTAR} observation. 
	The bottom panel shows the evolution of pulsed flux (12--25 keV) using {\it Fermi}/GBM. The vertical bars represent errors in corresponding measurements.}
	
\label{fig:freq}}
\end{figure}

Figure \ref{fig:freq} shows the evolution of the spin period and pulsed flux (12--25 keV) during the outburst using {\it Fermi}/GBM and the corresponding spin periods provided by {\it Fermi}/GBM are shown in the upper panel of the figure. The pulsed flux reached a maximum value of $\sim$0.24 keV cm$^{-2}$ s$^{-1}$. The top panel of this figure implies that the pulse period of the X-ray pulsar continuously decreased during the outburst. 

The evolution of pulse profiles is studied using different {\it NICER} observations at different luminosity levels during the outburst. Figure \ref{fig:profile_nicer} shows pulse profiles for different {\it NICER} observations. The pulse profiles do not show significant evolution during different {\it NICER} observations. The {\it NICER} profiles are more or less single peaks and nearly sinusoidal during  observations.

We look for any correlation between the spin-up rate and luminosity for the source. We have used the values of spin-frequencies during the outburst, which are provided by the \textit{Fermi}/GBM team. We utilized 18 spin frequency ($\nu$) measurements made during our study.  The spin frequency measurements were taken at equal  intervals of about 4 days. Each of the three subsequent frequency measurements with time was fitted using a linear function. The spin-up rate ($\dot \nu$) was determined from the slope of the linear function during a 12-day interval using the $\chi^2$ minimization method. The method was then repeated for the following three frequency measurements, and so on, \cite{Ka20}. As a result, from 18 spin frequencies, we obtained 6 spin-up rates. To obtain good fitting results we have also included 21 frequency measurements from the 2015 outburst, which provided additional 7 spin-up rates.

We have used the {\it NuSTAR} data to estimate unabsorbed flux (3--79 keV) during the {\it NuSTAR} observation of the 2021 outburst. We calculate the conversion factor by which the \textit{GBM} pulsed flux is related to the total flux from the {\it NuSTAR} observation. We multiplied all \textit{GBM} fluxes with the estimated conversion factor to convert them into the total flux. We have used the average value of total flux for three consecutive points over the same intervals, which are used to determine $\dot{\nu}$. Finally, the luminosity is estimated from the X-ray flux for a source distance of $\sim $20 kpc \citep{Ts16}. The X-ray luminosity of the source is estimated from the {\it Fermi}/GBM pulsed flux provided by \textit{GBM} team by multiplying a flux conversion factor of $6.85\times 10^{-9}$~erg\,keV$^{-1}$. 

The variation of spin-up rate with luminosity is shown in Figure~\ref{fig:GL}. The variation of spin-up rate with luminosity is fitted using a power law, which indicates the spin-up rate and luminosity are positively correlated. The spin frequency derivatives vary between $\sim $(0.4--2.0)$\times 10^{-11}\text{ Hz}\,\text{s}^{-1}$, which is estimated from spin frequency evolution history as provided by \textit{Fermi}/GBM. The luminosity is varied between $\sim $(3--8)$\times 10^{37}$~erg\,s$^{-1}$, which is estimated from the \textit{Fermi}/GBM pulsed flux using a multiplying factor.

\begin{table*}[t]
\centering
	\caption{Spectral fitting parameters for best fitted models for different {\it NICER} observations:}

\label{tab:Spec}
	\begin{tabular}{llcccc} \\
\hline

	Model   & Parameters                & {\it NICER} Obs 1         & {\it NICER} Obs 2        &  {\it NICER} Obs 3      &   {\it NICER} Obs 4      \\
\hline

               & $N_{H}$ & $3.0\pm0.3$  & $3.2\pm0.3$   & $3.8\pm0.2$  & $3.5\pm0.2$    \\
{\tt phabs$\times$(po+bb}) &$\Gamma$   & $0.62^{+0.3}_{-0.4}$  & $0.67^{+0.24}_{-0.29}$   & $1.08\pm0.08$ & $1.0\pm0.1$    \\
             & $kT_{bb}$ (keV)                 & $1.2^{+0.3}_{-0.12}$  & $1.0^{+0.12}_{-0.14}$   & -- & --  \\
             &   norm$_{bb}$, 10$^{-3}$        & $1.23\pm0.6$  & $0.6\pm0.4$   & -- & --  \\	
        	& Reduced $\chi{^2}$  (d.o.f)             & 1.13 (606)   & 1.07 (626)    & 1.02 (220) & 0.99 (174)   \\
              & Total Flux (0.6--10 keV)                & 3.55        & 3.04         & 2.87   & 2.63        \\
		&  &          &               &              &                  \\

\hline
\end{tabular}
	\begin{tablenotes}
		\small
	\item  $N_{H}$: hydrogen column density (10$^{22}$, cm$^{-2}$), $\Gamma$: power-law photon index, $kT_{bb}$: blackbody temperature, flux: (10$^{-10}$, erg cm$^{-2}$ s$^{-1}$). All of the reported errors are at 1$\sigma$ c.l., obtained using the {\tt err} tool from {\tt XSPEC}.
	\end{tablenotes}
\end{table*}

\begin{figure}[t]
\centering{
\includegraphics[width=6.2cm,angle=270]{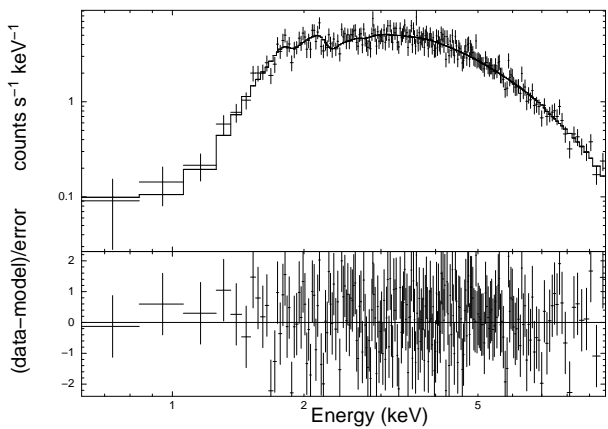}
\caption{The figure shows the {\it NICER} energy spectrum for obs 3, fitted with phabs$\times$(power law). Residuals are shown in the bottom panels of the figure.}
\label{fig:spectrum}
	}
\end{figure}

\subsection{Phase average energy spectra}
We have studied the energy spectra for different days of the outburst using {\it NICER} observations and look for the variation of different spectral parameters. The energy spectra with the best-fitted model are shown in Figure \ref{fig:spectrum} where the bottom panels of Figure \ref{fig:spectrum} show the residuals. We have extracted the energy spectrum for different observations from {\it NICER} and fitted them in {\tt XSPEC} with varying the model parameters independently for different models. 
The energy spectrum of the X-ray pulsar can be well fitted with a power-law ({\tt power law} in {\tt XSPEC}) and a blackbody emission component ({\tt bbody} in {\tt XSPEC}) along with photoelectric absorption ({\tt phabs} in {\tt XSPEC}). The spectra in the energy range of 0.6--10 keV is well described by blackbody emission with temperature (kT$_{bb}$) $\sim$0.9--1.3 keV and a neutral absorption of an equivalent hydrogen column density $\sim$3.5$\times$10$^{22}$ cm$^{-2}$. Table \ref{tab:Spec} shows the evolution of different spectral parameters with the best-fitted values for different days of the outburst. The photon index varied between $\sim$(0.2 to 1.1) during different {\it NICER} observations.  For the {\it NICER} observations 3 and 4, the spectra are fitted well with a power law component along with a photoelectric absorption. The additional black body component is not necessary to fit these two {\it NICER} spectra.

\subsection{Luminosity dependence of pulsed fraction and photon index}

\begin{figure}
\centering
\includegraphics[width=8.5cm]{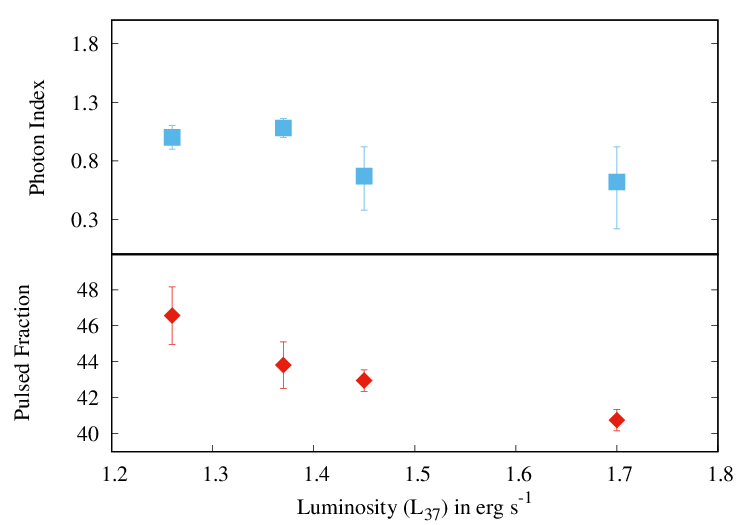}
\caption{Evolution of photon index and pulsed fraction with luminosity. Luminosity is estimated from the NICER X-ray flux in the energy range of 0.6--10 keV for a source distance of 20 kpc.}
\label{fig:evospec}
\end{figure}

We have looked for the variation of photon index and pulsed fraction with  luminosity during the outburst. 
The top panel of Figure \ref{fig:evospec} shows the variation of the photon index with luminosity. The photon index is estimated from the power-law continuum of energy spectra. The top panel of Figure \ref{fig:evospec} shows a weak anti-correlation between the photon index and luminosity. The bottom panel of Figure \ref{fig:evospec} shows the evolution of PF with luminosity using {\it NICER} observations. A negative correlation between luminosity and pulse fraction is observed during {\it NICER} observations below the luminosity of $1.8\times10^{37}$ erg s$^{-1}$. The pulsed fraction varies between 40\% and 48\% as estimated using {\it NICER} observations. The luminosity varies between $(1.25-1.7)\times10^{37}$ erg s$^{-1}$ during the {\it NICER} observation. The luminosity is estimated for a source distance of 20 kpc from {\it NICER} flux in the energy range of 0.6–10 keV.
\section{Discussion}
\label{dis}
We present the results from timing and spectral analysis of 2S 1553--542 during the recent outburst in January-February 2021 using the {\it NuSTAR}, and {\it NICER} data. 
The critical luminosity for 2S 1553--542 is estimated to be $\simeq$3.7$ \times$ 10$^{37}$ erg s$^{-1}$ by \citet{Be12}.  
The energy dependence of the pulse profile is studied to investigate the evolution of the individual peaks and beaming pattern of the pulse profile of the pulsar with different energies. The pulse profile shows a single peak and a nearly sinusoidal nature. 
In the energy range of 25–30 keV, the pulse profile shows an additional feature that is close to the line energy. The pulse fraction shows a trend to increase with energy, which is typical for an X-ray pulsar (viz. \citep{Lu09}). The pulse fraction shows a local feature near the cyclotron line energy during the {\it NuSTAR} observation. We  study the temporal variability of the pulse profile and pulse fraction during the outburst using {\it NuSTAR} and {\it NICER} observations during the outburst. Earlier, the pulse fraction varied with time during the outburst and decreased gradually with the outburst as observed by \citet{Pa12}. 

The beaming pattern of a pulsar is affected by its luminosity. When the source luminosity goes below the critical luminosity (subcritical regime), accreting material falls directly on the surface of the neutron star, producing a `pencil beam' X-ray emission. Emission escapes from the top of the column for the pencil beam pattern \citep{Bu91}. However, the beaming patterns can be more complex than a simple pencil or fan beam \citep{Kr95, Be12}. The pulse profile of the pulsar 2S 1553–542 during the 2021 outburst is a nearly single peak feature. These patterns are mostly governed by the ``pencil beam'' form. 

{\it Critical luminosity ($L_\textrm{crit}$) is important to define two accretion regimes of a source. In the subcritical regime, the source luminosity is lower than the critical luminosity, and at the critical luminosity, a state transition occurs from the subcritical to the supercritical regime. 
The radiation pressure in the supercritical regime is high enough to stop accreting matter at a distance above the neutron star, generating a radiation-dominated shock \citep{Ba76, Be12}. In the subcritical regime, accreted material reaches the surface of a neutron star through nuclear collisions with atmospheric protons or Coulomb collisions with thermal electrons \citep{Ha94}. Changes in cyclotron line energy, pulse profiles, and spectral shape can be used to probe these accretion regimes \citep{Re13, Pa89}.}

Earlier, near the cyclotron line energy, a distinct change in pulse profile and pulsed fraction was reported for a few sources, such as 1A 1118--61, Her X-1, GX 301--2, and 1A 0535+262 \citep{Ma22}.  Earlier, \cite{Lu09} summarized different sources for which a distinct change in pulse profile and pulse fraction was observed. A non-uniform increase in pulsed fraction with energy was observed for different sources like V 0332+53, Vela X-1, Her X-1, GX 1+4, 1A 0535+262, and 4U 0115+63 and has local maxima near cyclotron line energy and harmonics was found \citep{Lu09}.

For a few sources, the pulsed fraction exhibits a transition above the critical luminosity. In the supercritical accretion regime, the pulsed fraction shows a negative correlation with luminosity.  During the 2020 giant outburst, 1A 0535+262 showed a positive correlation between the PF and luminosity in low luminosity and above the critical luminosity, the correlation was reversed to a negative correlation \citep{Ma22}. We report an anti-correlation between the PF and luminosity for 2S 15553--542 during the 2021 outburst. 
Above the critical luminosity, the impact from the unpulsed photons leaving through the side walls of the column increases, which may affect the beaming patterns of the pulsar.

The 0.6--10 keV {\it NICER} spectra are fitted with a power-law and blackbody along with a photoelectric absorption model. Earlier, the X-ray spectrum of the source using XRT (0.5--10 keV) was modelled with simple model components like power-law or a blackbody emission along with photoelectric absorption \citep{Lu16}. Earlier, an iron emission line near 6.4 keV was reported from {\it RXTE} and {\it NuSTAR} observations \citep{Pa12, Ts16}. We have added a Gaussian component to the model to check the significance of the iron emission line, and it did not improve the goodness of the fit of {\it NICER} spectra. The evolution of different spectral parameters is studied during the outburst, which suggests that the spectra from different {\it NICER} observations can be explained with a  simple model like power-law  and blackbody along with photoelectric absorption. 

A negative correlation between the photon index and {\it NICER} flux is observed. The photon index decreases with an increase in X-ray flux, which indicates a ‘harder when brighter’ trend.
In the subcritical accretion regime, the negative correlation implies the hardening of the power-law continuum with flux. 
Earlier, several sources showed a significant variation in the L~--$~\Gamma$ diagram close to critical luminosity. The transition from a negative to positive correlation was seen in the L~--$~\Gamma$ diagram as luminosity increased \citep{Re13}. 
Earlier, in the subcritical regime, a negative correlation was reported for the sources like 1A 1118--612, GRO J1008--57, XTE J0658--073, and a transition in the correlation of $L$~--$~\Gamma $ was observed for the sources 1A 0535+262 \citep {Ma22}, 4U 0115+63, EXO 2030+375 \citep{Ep17, Ja21}, 2S 1417--624 \citep{Se22}, and KS 1947+300 \citep{Re13}.

\begin{figure}[t]
\centering{
\includegraphics[width=8.8cm]{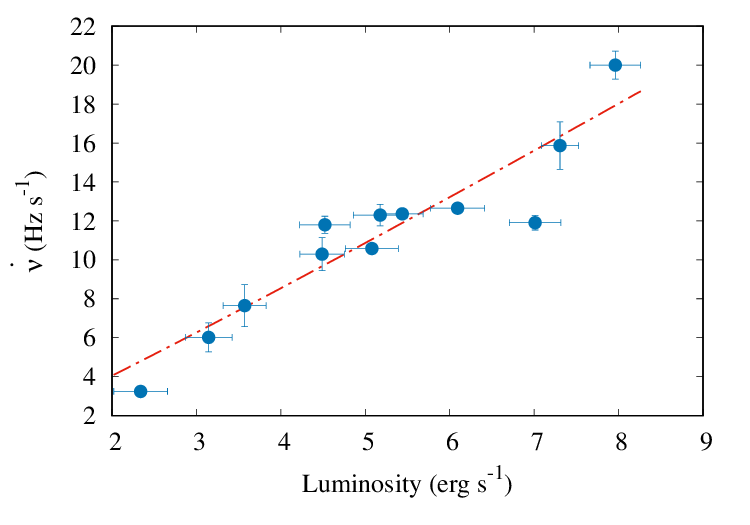} 
	\caption{ Variation of spin change rate (in unit of $10^{-12} \text{Hz}\,\text{s}^{-1}$) with luminosity (in unit of $10^{37}$~erg\,s$^{-1}$). The dotted blue line represents the best power-law fit of data points that gives a power-law index of 1.08$\pm $0.17.}
\label{fig:GL}}
\end{figure}

 Figure~\ref{fig:GL} shows that the pulse frequency derivatives of the X-ray pulsar 2S 1553--542 are correlated with luminosity. Earlier, during outbursts for several transient systems, a correlation between spin-up rate and X-ray flux was seen. This correlation was explained in terms of accretion. For example, GRO J1744--28 \citep{Bi97}, 2S 1417--624 \citep{Fi96a, Ma22b}, A 0535+26 \citep{Fi96b, Bi97}, EXO 2030+375 \citep{Pa89, Re96}, and SAX J2103.5+4545 \citep{Ba02} showed correlation between spin-up rate and X-ray flux. 

We have tried to calculate the magnetic dipole moment and the surface magnetic field of 2S 1553--542 using the accreting torque model and the observed spin-up rate. Transient X-ray pulsars are known to have a correlation between the spin-up rate and luminosity as \citep{Gh79b, Su17}:
%
\begin{equation}
\dot{\nu}_{12}=2.0 n \zeta ^{\frac{1}{2}} \mu _{30}^{\frac{2}{7}} R_{6}^{
\frac{6}{7}} M_{1.4}^{-\frac{3}{7}} I_{45}^{-1} L_{37}^{\frac{6}{7}}
\end{equation}
where $\mu _{30}$, $\dot{\nu}_{12}$, $R_{6}$, $M_{1.4}$, and $I_{45}$ are the  magnetic dipole moment, spin frequency derivative, radius, mass, and the moment of inertia of the neutron star given in the units of $10^{-12}\text{ Hz}\,\text{s}^{-1}$, $10^{30}$~G\,cm$^{3}$, $10^{6}$~cm, 1.4~$M_{\odot}$, and $10^{45}\text{ g}\,\text{cm}^{2}$ respectively. $L$ is the X-ray luminosity in the unit of $10^{37}$~erg\,s$^{-1}$. According to the \citet{Gh79a,Gh79b} model, under slow-rotator condition, $n\sim $1.39 and $\zeta \sim 0.52$. Therefore, equation (1) reduces to \citet{Su17}
%
\begin{equation}
\dot{\nu}_{12}=k L_{37}^{\alpha}
\end{equation}
where $k = 2.0 \mu _{30}^{\frac{2}{7}}$ and $\alpha  =
\frac{6}{7}$. For the nominal values of $I_{45} = R_{6} = M_{1.4} = 1$, measurements of the $\dot{\nu}$ versus $L$ give a rough estimation of the magnetic dipole moment of the pulsar. From the $L$ vs $\dot{\nu}$ plot, we have calculated $k$ and $\alpha $ as 1.92$\pm $0.58 and 1.08$\pm $0.17 respectively using the best-fit results. The estimated value of $\alpha $ is close to the theoretical value. 
Figure~\ref{fig:GL} shows the correlation between the spin-up rate and luminosity and the dotted line represents the best-fitted result. From the best-fit result, we may write equation (2) as
%
\begin{equation}
\dot{\nu}_{12}=(1.92\pm 0.58) ~L_{37}^{1.08\pm 0.17}
\end{equation}
The magnetic dipole moment of the pulsar can be written in the form
%
\begin{equation}
2.0 ~\mu _{30}^{\frac{2}{7}} = 1.92\pm0.58 ; \mu _{30} \simeq 0.87\pm0.01
\end{equation}

The surface magnetic field can be calculated from the magnetic moment ($\mu _{30}$) and radius ($R_{6}$) of the pulsar as
%
\begin{equation}
\mu _{30} = \frac{1}{2} B_{12}R_{6}^{3}\phi (x)
\end{equation}
$\phi (x)$ is the correlation factor, for typical NS, $\phi (x)$
$\sim $0.68. The magnetic field can be written as
%
\begin{equation}
B_{12} = 2 \times \frac{\mu _{30}}{0.68}
\end{equation}
for $\mu _{30}$ $\simeq $~0.87, the magnetic field is estimated to be
$\simeq  (2.56\pm0.03) \times  10^{12}$~G.

We have also estimated the magnetic field corresponding to the critical luminosity. Earlier, the critical luminosity was reported to be $4\times 10^{37}$~erg\,s$^{-1}$ \citep{Mal22}.
For a typical neutron star, the magnetic field is related to the critical luminosity as \citet{Be12},
%
\begin{equation}
L_{\mathrm{critical}} = 1.5 \times 10^{37}\left (\frac{B}{10^{12} G}
\right )^{\frac{16}{15}}~\text{erg}\,\text{s}^{-1}
\end{equation}
The magnetic field corresponding to the critical luminosity $4\times 10^{37}$~erg\,s$^{-1}$ is estimated to be $2.5 \times 10^{12}$~G for a source distance of 20 kpc. This value of the estimated magnetic field closely matches our results.

\section{Conclusion}
\label{con}
We have summarized the results of the timing and spectral analysis of the poorly studied X-ray pulsar during the outburst in 2021. The spin period of the neutron star was found to be $P \sim$ 9.2822 s based on {\it NuSTAR} data. The pulse profile using {\it NuSTAR} shows a single peak feature, and near the cyclotron line energy, an additional feature in the pulse profile is observed.
The beaming patterns were mostly dominated by pencil beams during the {\it NICER} and {\it NuSTAR} observations. The pulse fraction positively correlates with energy with a local feature near the cyclotron line energy. The pulse fraction shows a negative correlation with luminosity ($Lx<1.8\times10^{37}$ erg s$^{-1}$) for {\it NICER} observations. The source enters the supercritical regime at the highest flux above the critical luminosity. The spectrum can be described using an absorbed power law with a blackbody component. A variation of the photon index with luminosity is studied, which implies that the photon index decreases with an increase in luminosity during NICER observations. The negative correlation implies that the source was in the sub-critical accretion regime during those observations. The variation of spin-up rate with luminosity is studied for 2S 1553--542 during the outburst, which suggests the spin-up rate is strongly correlated with luminosity. The magnetic field is estimated using the torque-luminosity model with different spin-up rates. The magnetic field is calculated to be $ 2.56  \times 10^{12}$ G from the torque-luminosity model for a source distance of 20 kpc.

\section*{Acknowledgements}
 We thank the anonymous reviewer for his/her suggestions, which help to improve the manuscript significantly.
This research has been done using data collected by {\it NuSTAR}, a project led by Caltech,  managed by NASA/JPL and funded by NASA, and has utilized the {\tt NUSTARDAS} software package, jointly developed by the ASDC (Italy) and Caltech (USA). This research has made use of the {\it MAXI} data provided by the RIKEN, JAXA, and {\it MAXI} team. We acknowledge the use of public data from the {\it NuSTAR}, {\it NICER}, and {\it Fermi} data archives.

\section*{Data availability}
The data underlying this article are publicly available in the High Energy Astrophysics Science Archive Research Center (HEASARC) at \\
\url{https://heasarc.gsfc.nasa.gov/db-perl/W3Browse/w3browse.pl}

\section*{Conflict of interest}
The authors declare that they have no known competing financial interests or personal relationships that could have appeared to influence the work reported in this paper.

\section*{Author contribution}
Mr. Mandal took part in the analysis of data, the visualization of results, and manuscript writing. Dr. Pal contributed to the conceptualization of the project and the manuscript review and writing.

{}

\appendix

\section{Pulse profile evolution from {\it NICER} observations}
\begin{figure*}[t]
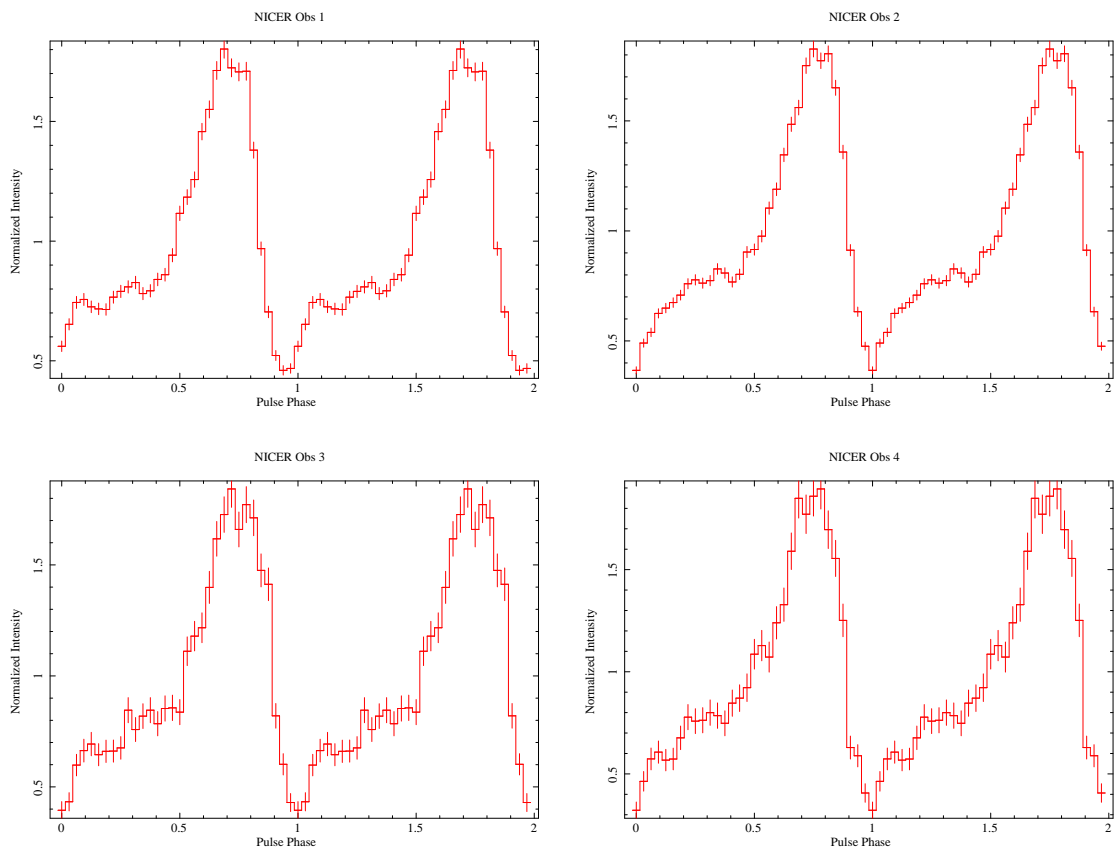

\centering{
\includegraphics[width=5.8cm,angle=270]{profile_ni101.eps}
\includegraphics[width=5.8cm,angle=270]{profile_ni102.eps}
\includegraphics[width=5.8cm,angle=270]{profile_ni103.eps}
\includegraphics[width=5.8cm,angle=270]{profile_ni104.eps}
\caption{Pulse profiles of 2S 15553--542 for different {\it NICER} observations during the outburst. The top left shows the pulse profile for {\it NICER} Obs 1, the top right is for Obs 2, the bottom left shows the pulse profile for Obs 3 and the bottom right is the pulse profile for Obs 4.}
\label{fig:profile_nicer}
}
\end{figure*}

Figure \ref{fig:profile_nicer} shows the pulse profiles of 2S 1553--542 during different {\it NICER} observations.

\end{document}